\documentclass[journal, compsoc]{IEEEtran}
\usepackage{amsmath,amsfonts}
\usepackage{algorithmic}
\usepackage{algorithm}
\usepackage{array}
\usepackage[caption=false,font=normalsize,labelfont=sf,textfont=sf]{subfig}
\usepackage{textcomp}
\usepackage{stfloats}
\usepackage{url}
\usepackage{verbatim}
\usepackage{graphicx}
\usepackage{cite}
\usepackage{multirow}
\usepackage{pifont}
\usepackage[inline]{enumitem}

\usepackage[dvipsnames, table,xcdraw]{xcolor}
\usepackage{hyperref}
\usepackage{cleveref}
\hypersetup{
  linkcolor  = Aquamarine,
  citecolor  = Aquamarine,
  urlcolor   = Aquamarine,
  colorlinks = true,
}
\usepackage{colortbl}
\usepackage{adjustbox}
\usepackage{soul}

\reversemarginpar
\begin{document}

\title{The Emotion-Memory Link: Do Memorability Annotations Matter for Intelligent Systems?} 

\author{Maria Tsfasman, Ramin Ghorbani, Catholijn M. Jonker, Bernd Dudzik \vspace{-1em}}

\markboth{Journal of \LaTeX\ Class Files,~Vol.~14, No.~8, August~2021}%
{Shell \MakeLowercase{\textit{et al.}}: A Sample Article Using IEEEtran.cls for IEEE Journals}

\IEEEpubid{0000--0000/00\$00.00~\copyright~2021 IEEE}

\IEEEtitleabstractindextext{%
\begin{abstract}
Humans have a selective memory, remembering relevant episodes and forgetting the less relevant information. Possessing awareness of event memorability for a user could help intelligent systems in more accurate user modelling, especially for such applications as meeting support systems, memory augmentation, and meeting summarisation. Emotion recognition has been widely studied, since emotions are thought to signal moments of high personal relevance to users. The emotional experience of situations and their memorability have traditionally been considered to be closely tied to one another: moments that are experienced as highly emotional are considered to also be highly memorable. This relationship suggests that emotional annotations could serve as proxies for memorability. However, existing emotion recognition systems rely heavily on third-party annotations, which may not accurately represent the first-person experience of emotional relevance and memorability. This is why, in this study, we empirically examine the relationship between perceived group emotions (Pleasure-Arousal) and group memorability in the context of conversational interactions. Our investigation involves continuous time-based annotations of both emotions and memorability in dynamic, unstructured group settings, approximating conditions of real-world conversational AI applications such as online meeting support systems. Our results show that the observed relationship between affect and memorability annotations cannot be reliably distinguished from what might be expected under random chance. We discuss the implications of this surprising finding for the development and applications of Affective Computing technology. In addition, we contextualise our findings in broader discourses in the Affective Computing and point out important targets for future research efforts.
\vspace{-1em}
\end{abstract}

\begin{IEEEkeywords}
conversational memory; affect; social signal processing; memory encoding; memory retention; group interaction; relevance
 \vspace{-1em}
\end{IEEEkeywords}}

\maketitle
\IEEEpeerreviewmaketitle
\IEEEdisplaynontitleabstractindextext

\section{Introduction\label{sec: introduction}}

Memory for conversations and other social interactions plays a crucial role in shaping social bonds and fostering relationship building \cite{Bluck2005FunctionsMemory}. Considering human conversational memory in intelligent systems is, thus, essential for explaining and predicting human behaviour in conversations including their affective responses. Conversational memory can be defined as a sub-type of episodic memory, which manages the encoding, storage, and retrieval of personally experienced events \cite{NeuroEncyclopedia2021}, particularly within conversational settings. 

Affective Computing (AC) has long focused on recognising and interpreting human emotions to enhance interactions between users and intelligent systems \cite{poria2017AffectiveComputingReview}. Emotions are considered to be central to human experience, shaping decision-making, social interactions, and memory. Their automatic detection is valuable for intelligent systems because emotional responses often signal moments of high personal relevance to users. To capture these signals, Multimodal Emotion Recognition (MER) commonly uses human behavioural cues, such as facial expressions, speech patterns, and physiological signals, to infer emotional states. While MER has made significant strides in detecting momentary affective states, its potential to model longer-term cognitive processes, such as memory, remains under-explored.

Both theoretical and empirical research suggests that the way we emotionally experience events is strongly linked to how well we remember them \cite{mcgaugh2013}. Emotional arousal enhances memory encoding and retrieval, with emotionally charged events being remembered better than neutral ones \cite{Dolcos_2004_EmotionAmygdaleMemory}. The effect is linked to hormone release during arousing experiences, which strengthens memory formation in the brain \cite{Cahill_1996_MemoryModulation}. Both valence \cite{Maljkovic_Martini_2005ValenceAffectsMemory, Erk_2003EmotionalContextAndMemory} and arousal \cite{Sharot2004, Cahill_1996_MemoryModulation}, as well as their combination \cite{Gomes_Brainerd_Stein_2013ValenceArousalMemory}, have been shown to enhance memory processes. Additionally, affect and memory are closely tied to the personal relevance of a stimulus, as relevance influences both emotional experience \cite{Olteanu_2019_RelevanceModulatesEmotion} and the likelihood of remembering an event \cite{Shohamy_2010_MotRelevenceInfMemory}. Given these well-documented relationships, it is reasonable to hypothesise that perceived affect could serve as a proxy for memory. 

These and similar findings have been used to motivate various intelligent systems to integrate emotional components into computational memory models, e.g., to drive interactions between users and virtual agents \cite{ kasap_interacting_2010, Martin2021AffectiveMemory, brom_towards_2009}, social robots \cite{kasap_towards_2010, Ahmad2021EmotionMemoryModel}, or between agents in multi-agent systems \cite{Moerland2017EmotionForRL}. 

However, despite this prevalent conceptual connection between emotional responses and memorability, Affective Computing research has not yet explored this connection in the context of MER technology. Unfortunately, without targeted empirical exploration, it remains unclear to what extent theory and findings from the behavioural sciences connecting the two phenomena actually translate to many of the settings in which MER technology is developed or expected to operate. Notably, the following aspects span crucial practices of the development and deployment of MER technology but are not sufficiently covered in existing findings connecting the two phenomena:

\begin{enumerate}
    \item \textbf{Choice of Annotation Perspectives:} Although research robustly links experienced emotions (measured through self-reports and physiological signals) to memory encoding, it remains uncertain whether third-party observed affect annotations, which are widely used in affective computing, can reliably serve as proxies for personal memorability (more detail in Section \ref{sec:study_object}).
    \item \textbf{Continuous Conceptualisation}: Previous research on the link between affect and memory operationalises those as static states, while in MER systems it is more common and desirable to view those as continuous. The link between continuous annotation of affect and memory has not been studied, to our knowledge
    (see Section \ref{sec:continous}).
    \item \textbf{Group-based Analysis}: Group-based MER systems are crucial for real-world applications like meetings and collaborative tasks, yet existing research on the emotion-memory link primarily focuses on individuals, overlooking the social dynamics that shape both affect and memory in group interactions (see Section \ref{sec:group-level}).
\end{enumerate}

Given these gaps, the extent to which third-party affect annotations capture memory-relevant information in real-world settings remains an open question. To address this question, in this paper, we present an empirical investigation evaluating the association between annotations of Perceived Group Emotions (Pleasure-Arousal) and Group Memorability. Our study leverages time-continuous annotations of emotions and memorability in dynamic, unstructured group interactions, mirroring real-world conditions relevant to MER applications like online meeting support. We discuss the implications for Affective Computing, situate our findings within broader Affective Science discussions, and highlight key directions for future research to bridge the gap between computational modelling of affect and memory.

\section{Background and Motivation}

\begin{figure*}
    \centering
    \includegraphics[width=\linewidth]{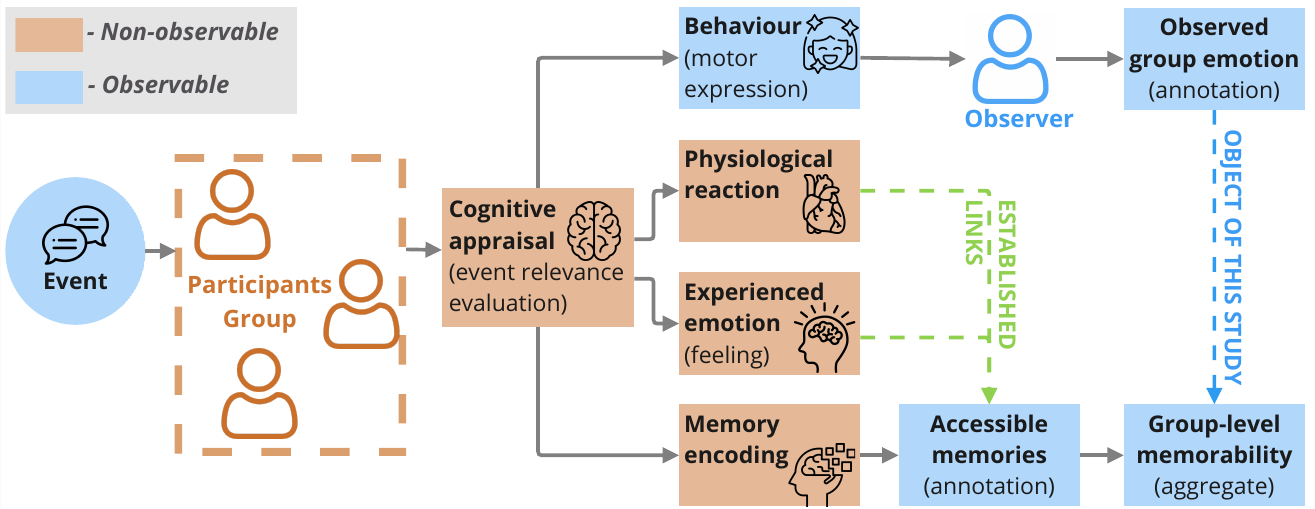}
    \caption{An illustration of perceived vs experienced emotion and the object of this study (shown with blue dashed arrow). The states with observable information are shown in blue and the states we have no access to are shown in orange. The green dashed arrow represents the relationships that have been studied in previous literature. (Based on the component model of emotions according to Scherer (1984) \cite{Scherer_1984})}
    \label{fig:study_object}
\end{figure*}

In this section, we briefly expand on some of the core properties of Affective Computing development practices that we believe could limit the insights that existing empirical findings on the emotion-memorability link can provide for the development and applicability of \textit{Multimodal Emotion Recognition (MER)} technology. To support this discussion, we provide a graphical overview of the components and relationships involved in the development practices we discuss in \textit{Figure \ref{fig:study_object}}.

\subsection{Choice of Annotation Perspective \label{sec:study_object}} 

As mentioned in Section \ref{sec: introduction}, there is substantial empirical evidence indicating a link between emotional episodes and the memorability of events. However, the findings supporting this link are largely grounded in physiological signals of emotions and self-reports of first-person experience (e.g. \cite{Sharot2004, Cahill_1996_MemoryModulation, Maljkovic_Martini_2005ValenceAffectsMemory, Erk_2003EmotionalContextAndMemory, Dolcos_2004_EmotionAmygdaleMemory}), i.e., they capture what we could call "Experienced Emotions" and "Self-reported Memorability" of events. When a person encounters an event, the event is thought to be experienced through a process of cognitive appraisal - evaluation of event relevance, implication, coping potential and normative significance \cite{Lewis2005Appraisal}. Such appraisal is thought to result in motor expression (behaviour), physiological reaction and a subjective feeling (experienced emotion) according to widely accepted Scherer's component model of emotion \cite{Scherer_1984}. Along with that, a process that is triggered is memory encoding (if the event is considered relevant). The well-established link between emotion and memory described above applies to the relationship between memory encoding and measures of experienced emotion (subjective feeling) \cite{Shohamy_2010_MotRelevenceInfMemory, Gomes_Brainerd_Stein_2013ValenceArousalMemory, Maljkovic_Martini_2005ValenceAffectsMemory, Erk_2003EmotionalContextAndMemory} or the physiological responses related to emotions \cite{Dolcos_2004_EmotionAmygdaleMemory, Cahill_1996_MemoryModulation} (shown with a green dashed arrow in Figure \ref{fig:study_object}) \footnote[1]{Note that these studies have focused on individual memory and emotion, not in group context}.

In the context of affective computing, however, it is more common to focus on observed emotion - a hired annotator (observer in Figure \ref{fig:study_object}) watches videos and labels arousal and valence of participants' emotional behaviour \cite{AffectDetection2015}. Relying on observed annotations is done on the one side, for pragmatic reasons - it is easier to employ external annotators than collecting self-reports, it is possible to hire many annotators for each temporal segment to reduce annotation subjectivity. On the other side, it is more practical for modelling reasons, since annotators rely on external behaviour to make their judgement, similarly to what an emotion recognition model would do.

This distinction between experienced and observed emotion is important since the observed emotion is an external evaluation of a participant's behaviour by a third-party observer. Since such third-party annotation is solely based on the observed behaviour it may contain some inaccuracies. For example, not all emotions might be expressed through behaviour or sometimes the expressed behaviour might not reflect the experienced emotion because of social norms (e.g. covering anger with a polite smile to avoid confrontation). In addition, in the case of group affect \cite{Navin2024AffectMemo}, the third-party annotation reflects the observed emotions of the group rather than the individual participants one-by-one and might not be equal to specific emotions of each group member (e.g. if one participant displays a negative valence while three display highly positive valence, the valence of the group would probably be labelled as high).

In conclusion, while the literature points towards a clear physiological link between experienced affect and memory encoding, the question of whether observed affect annotations are representative of memory labels remains open. It remains uncertain whether previous insights from the behavioural sciences can inform research about the ability of Multimodal Emotion Recognition (MER) to provide insights into the memorability of situations.


\subsection{Time-Continuous Operationalisation of Concepts\label{sec:continous}}
A common practice in developing MER systems is to collect data that operationalises emotional responses through \textit{time-continuous measurements} (e.g., annotations collected for every frame in a video stream \cite{mariooryad2015a}). Some of the proposed benefits \cite{zhang2020a} of this practice are considered to be its high temporal granularity (i.e., being able to capturing nuanced changes in emotional qualities over time), but also its capacity to capture emotional variability (i.e., being able to describe changes in emotional qualities within some specified unit of analysis, such as a video clip). Pragmatically, it seems also plausible that time-continuous estimates of emotion are highly desirable from the point-of-view of many applications, since they might enable systems to respond to changes observed in users dynamically.  

While common in data collection for MER development, studies investigating the emotion-memorability link in the behavioural sciences do not operationalise either concepts with such time-continuous measures. Instead, these concepts are typically operationalised with self-reports describing events as reconstructed from memory, without access to information at encoding time or fine-grained breakdowns of parts of the event (e.g., see Talarico\&Rubin for an example \cite{Talarico2004}).
    
Given this misalignment in operationalisation, it seems unclear in how far existing findings about the emotion-memorability link can generalise to the outcome achieved with practices used for developing MER. For example, memory biases might distort how emotion is attributed to remembered events or emotional connections to memorability might be due to increased rehearsal over time \cite{mcgaugh2013} (see also Dudzik \& Broekens \cite{dudzik2023} for a more extensive discussion of potential influences manifested by the choice of when to provide emotional self-reports). Overall, it does not seem self-evident that estimates based on datasets that have a time-continuous affective ground truth reliably approximate memorability when similarly operationalised. 

For this reason, our study operationalises both concepts in a time-continuous way, leveraging the MeMo dataset \cite{Tsfasman2024} that contains relevant annotations for memorability, which were recently complemented with associated time-continuous Affect Annotations by Raj Prabhu et al. \cite{Navin2024AffectMemo}.

\subsection{Group-based Analysis\label{sec:group-level}} 
Many applications for which Multimodal Emotion Recognition is beneficial are often expected to operate in group settings, such as work meetings \cite{Samiha2021MeetingCoach}, educational settings \cite{Trian2019GroupAffectForLearning}, and collaborative tasks \cite{Järvenoja2017GroupEmotionsForCollaboration}. In these contexts, MER can not only be used to analyse individual emotions but also to assess group-level affective dynamics, where the target of predictions extends beyond individuals to an entire team and its emergent characteristics.

Group affect, the collective emotional state of a group, has gained attention in affective computing and computational modelling. It encompasses shared moods and emotions among group members during interactions \cite{Smith2007TrulyGroupAffect, Veltmeijer2023AutomaticERGroup}. Research has explored integrating group affect into decision-making processes, developing computational models that consider individual, group, and emerging processes \cite{chohra2018groupaffect}. Studies have investigated the dynamics of group affect, including convergence and divergence of affective expressions, using multimodal approaches to extract synchrony-based features from audio and visual cues \cite{Navin2024AffectMemo}.

Existing research on the relationship between emotion and memory in the behavioural sciences has primarily focused on individual experiences rather than group-based settings. Most studies investigating this link have been conducted in controlled environments where participants engage with stimuli in isolation, rather than within dynamic social interactions (e.g., \cite{Sharot2004, Cahill_1996_MemoryModulation, Maljkovic_Martini_2005ValenceAffectsMemory, Erk_2003EmotionalContextAndMemory, Dolcos_2004_EmotionAmygdaleMemory}). Social interactions such as conversations differ from individual contexts because of the continuous exchange of (non-)verbal signals and relational dynamics, changing the resulting quality and quantity of remembered information \cite{stafford_actor-observer_1989, benoit_participants_1996, mckinley_memory_2017, Samp2007_FriendsMemory, Knutsen2014EgoMemory, Miller_Interpersonal_2002, diachek_linguistic_2024}. Although there is some research on emotion and memory in social contexts, the literature linking the two concepts to each other  typically conceptualises them as personal cognitive-affective processes, without considering how these phenomena may emerge differently in conversations and other group contexts. This individual-level focus limits the applicability of prior findings to real-world scenarios where memory and emotions are often shaped by collective interactions.

This gap is particularly relevant for Affective Computing applications, as the cognitive-affective processes at play in group settings may differ significantly from those observed in isolated individuals. Social dynamics such as emotional contagion and regulation not only influence individual affective states but also the overall emotional climate of a group, potentially affecting how shared experiences are remembered, while also significantly impacting group interaction outcomes, including creativity, analytical performance, sense of belonging, and information sharing \cite{BarsadeGibson2012GroupAffect, Klep2011GroupAffect, Kim2020CohesionLearning}.  These factors suggest that findings from traditional emotion-memorability studies may have limited capacity to inform the development of MER systems intended for group-based applications, underscoring the need for research that explicitly addresses affect and memory at the group level.

For these reasons, we believe that our study's setup provides a meaningful addition to the existing body of research since it explicitly focuses on group-based analysis: it 1) takes place in group-conversational settings and 2) conceptualises both emotion and memorability as group-level constructs. 

\section{Related Work: Memory}

\subsection{Memorability prediction}  
The few studies that do investigate memorability from a computational perspective focus on memorability of media stimuli \cite{cohendet_videomem_2019, MemorabilityModeling2020, 2022MedievalEEGMem}. 
These studies have generally been successful in identifying features (such as semantic richness, emotional valence, and visual distinctiveness) that contribute to enhanced memorability of media segments across individuals \cite{cohendet_videomem_2019, MemorabilityModeling2020, 2022MedievalEEGMem}. Such memorability modelling has the potential to advance user modelling to understand what is actually relevant for the user and what needs to be repeated or reframed for a greater impact on the user in long-term. Nevertheless, in an arguably more common and socially important context of conversations, memory modelling remains underexplored. Unlike media memorability, where the stimulus is a fixed and repeatable entity, conversational memory emerges from dynamic, interactive, and multi-speaker contexts, making it more complex to model. Moreover, many conversations happen in a group of people, offering an additional complexity as well as a source of insights into what is considered memorable in a group (e.g. team meetings, friends and family gatherings).

\subsection{Conversational Memory Modelling} 

Although understanding the way humans remember conversational stimuli has been the subject of decades of cognitive research (see \cite{mckinley_memory_2017, benoit_participants_1996, Samp2007_FriendsMemory, Knutsen2014EgoMemory, Miller_Interpersonal_2002, diachek_linguistic_2024} for illustrative examples), it has only recently been approached from a computer science perspective. To our knowledge, only one study and dataset have addressed the task of predicting conversational memory: Tsfasman et al. 2022 \cite{Tsfasman2022} has introduced a baseline model using the MeMo conversational memory corpus \cite{Tsfasman2024}.

\section{Methods: Dataset}

\subsection{Data Source}
In this paper, we use data contained in the recently created MeMo dataset \cite{Tsfasman2024}. It was collected in an online video-conferencing setting of 45-minute longitudinal group conversations. Each group included 3 to 5 participants and one professional moderator. The moderator was tasked with keeping the conversation going and trying to keep the atmosphere in the group comfortable for participants to be ready to openly express their opinions and emotions. The topic of the conversations was the Covid-19 pandemic, focusing on people's experiences and opinions about the pre- and post-pandemic world (a relatable topic for many at the time of the recording of the corpus in 2021). Each group participated in three 45-minute long conversations scheduled 3-4 days apart. While the MeMo corpus originally does not contain any affect annotations, these were provided in a follow-up by Raj Prabhu et al. \cite{Navin2024AffectMemo}. 

\subsection{Data preparation}
In this paper, we used a subset of the original MeMo corpus: the sessions and specific timestamps for which Raj Prabhu et al. \cite{Navin2024AffectMemo} have collected affect annotations. This selection resulted in 3 groups from the original MeMo corpus being exploded and the timestamps at the start and end of the recordings being trimmed to match the timestamps of the annotations (for details on the exclusion criteria see the relevant publication \cite{Navin2024AffectMemo}). In addition, we excluded 4 sessions with gaps in memory annotation (i.e., instances where at least one participant did not fill out the post-session survey. The subset of MeMo used in this paper consisted of 30 conversational sessions totalling 1457 minutes of recording (mean of 41.6 min. per video +- 7.5 min.), 12 groups with 42 participants in total.

\subsection{Annotation Collection}

\textbf{Perceived Group Affect} 
\label{affect_labels}
Group affect annotations are present in 15-second intervals across two affect dimensions: arousal and valence, based on Russell’s circumplex model \cite{Russell_1980}. The annotations are on an ordinal scale (1-9), allowing annotators to express varying intensities of affective states in a continuous manner. The study employed 8 annotators with backgrounds in organisational psychology and prior experience in annotating social behaviours. Each annotator underwent a training designed to ensure consistency and reliability in their evaluations. This training emphasised the importance of focusing explicitly on observable emotional behaviours, ensuring that annotations were grounded in visible cues rather than inferred internal states. Each group interaction video was assessed by at least 6 annotators. Inter-annotator reliability was evaluated, revealing moderate agreement across both affect dimensions. For additional details see Raj Prabhu et al. \cite{Navin2024AffectMemo}.

\subsubsection{Remembered Moments}
\label{memory_labels}
\textbf{Memory annotation procedure.} The memory annotation procedure consisted of two stages. First, immediately after each conversational session, participants completed an open-ended free-recall task, reporting up to 10 moments they remember in their own words with as much detail as possible in a free form. This step ensured that memories reflected participants’ accessible recollections at the time (for more detail see \cite{Tsfasman2024}). Second, participants reviewed the session recordings to match their reported moments to specific events by providing start and end times or indicating if the memory lacked a precise interval. This self-assignment ensured memory-event alignment based on participants’ perspectives rather than third-party interpretations. The process was designed to prioritise construct validity while managing participant fatigue and minimising bias (see more details on MeMo setup in \cite{Tsfasman2024}). 

In total, there were 419 self-reported memorable moments in the data used in this paper, with a mean duration of 110 seconds (+- 116, from 1 to 580 seconds)\footnote{The minimum duration of a memorable moment is equal to 1 second because the videos were cropped based on the valid segments used in the curated dataset and affect annotations. A one-second moment is likely part of a longer event, with its start or end removed for varying reasons (see \cite{Tsfasman2024} and \cite{Navin2024AffectMemo} for details on video cropping).}.

\subsection{Processing and Derived Measures}
In this section, we briefly outline how we processed the annotations described above to arrive at operationalizations of \textit{Perceived Group Affect} and \textit{Group Memorability}. 

\subsubsection{Group Memorability Measures}
\textbf{Continuous Group Memorability Index.}
To compute the combined measure of what segments were memorable to the group, for each second of the recording we calculate a Group Memorability Index. It is defined as the ratio of participants from a group that included those timestamps in their annotations for Remembered Moments.

\textbf{Binary Group Memorability Index.}
From the continuous memory measure, we compute a boolean metric of whether each individual second was recalled by at least one member from within the group. The memory boolean value was 0 if the memory index was 0. If the memory index was greater than 0 the value was 1. In essence, this variable encodes the most generous operationalisation of the group memorability concept.  

\subsubsection{Perceived Group Affect Measures} 
\textbf{Continuous affect: Valence and Arousal} We aggregate the arousal and valence annotations across raters using a median of all the reported scores for each second. In addition to aggregation, we shift the scales to be able to compute an additional measure of affect intensity (see below).

Specifically, we shifted the values for both dimensions to a Likert scale from 0 to 8. For arousal we used the following formula:
\begin{equation}
    A'_t = A_t - 1
\end{equation}
where $t$ represents the timestamp (in seconds).

To better reflect the bipolar nature of valence, shifting and rescaled the original values:
\begin{equation}
    V'_t = 2 \cdot (V_t - 5)
\end{equation}

This changes the original 1-to-9 scale to a symmetric range from $-8$ to $8$, ensuring that the intensity of both negative and positive valence is represented with equal magnitude.

\textbf{Continuous affect: Intensity} 
Some versions of the emotion-memorability link conceptualize the emotional component not in terms of specific emotional qualities (e.g., pleasure), but instead as the intensity of the emotional episode at the time of encoding \cite{WOLF2021103119IntensityMatters}. According to Reisenzein \cite{Reisenzein_1994Intensity}, emotional intensity refers to the strength or magnitude of an emotional experience, which can be understood in terms of the circumplex model of affect \cite{Russell_1980}. This model posits that emotions are organised along two primary dimensions: pleasure (valence) and arousal. Emotional intensity represents the degree to which a person experiences these two dimensions, ranging from mild to strong. In this framework, the intensity of an emotion can be computed based on the values of arousal and valence \cite{Goto2017IntensityEncyclopedia}. In this paper, emotional intensity labels $\mathbf{I}$ for timestamp $\mathbf{t}$ is calculated as the Euclidean norm of valence score $V_t$ and arousal score $A_t$:
\begin{equation}
\label{intensity}
    \mathbf{I_t} = \sqrt{V_t^2 + A_t^2}
\end{equation}
The resulting intensity metric ranges from $0$ to $\sqrt{8^2+8^2}$($~$11.3).

\textbf{Affect - Binary} To investigate if the relationship between group affect and memory annotations is more clear with binary affect values with a meaningful threshold, we computed binary affect labels with the middle of each annotation scale as a threshold: 0 for valence, 4 for arousal and 4 for intensity. The threshold of 4 for intensity does not correspond to the mathematical midpoint of the intensity range, but is instead derived from a functionally meaningful point: the intensity calculated when arousal is at its midpoint (4) and valence is neutral (0). That is:

\begin{equation}
\mathbf{I_{mid}} = \sqrt{V_{mid}^2 + A_{mid}^2} = \sqrt{0^2 + 4^2} = 4
\end{equation}

This threshold reflects a moderate level of emotional engagement based on the meaningful input scales, rather than a purely statistical midpoint. It is thus used to distinguish between lower and higher affective intensity in a way that reflects the underlying annotation schema and preserves interpretability in relation to arousal and valence contributions.

\section{Methods: Analysis \label{evaluation}}

\subsection{Metrics}
Comparing human internal states to each other to evaluate their alignment brings several challenges. First, human data is typically characterised by noise and is prone to errors, especially when it comes to perceived measurements such as affect (e.g., there might be a delay in annotation \cite{mariooryad2015a} or misinterpretations or lapses in annotators' attention). Second, there might be confounding variables at play, such as the introduction of new stimuli or non-measured factors at play. Third, when dealing with continuous data such as human interactions and changing internal states, there is temporal context that needs to be taken into account. To address these challenges, it is essential to use evaluation metrics that are sensitive to the temporal nature of the data. Traditional measures of time-series comparison, such as Mean Squared Error (MSE) or Mean Absolute Error (MAE), penalise slight temporal shifts or discrepancies too harshly.  

\subsubsection{PATE}
Therefore, the first metric we rely on is a recently introduced metric designed specifically for time-series predictions that accounts for such temporal shifts - the \textsc{Proximity-Aware Time series Evaluation (PATE)} metric \cite{ramin2024pate}. PATE is conceptually similar to the Area Under the Precision-Recall Curve (AUCPR) but introduces additional considerations for time series. Specifically, PATE does not treat all mismatches between predictions and ground truth equally; instead, it introduces a tolerance window around each ground truth event, recognising that in real-world human-centred data, small temporal shifts in responses (such as memory recall or affect expression) are expected. The metric assigns partial credit to predictions that fall within this window, reducing the penalty for minor misalignments. The PATE framework introduces the assumption that temporally proximate events are functionally related, making it particularly well-suited for evaluating human behavioural data where slight timing discrepancies should not be treated as errors.

In this paper, we use the standard \textbf{PATE} measure, which operates on one binary time-series (memory boolean in our case) and one continuous (affect annotations), by setting various thresholds on the continuous data and computing the area under the curve as the output PATE value. Additionally, we use \textbf{PATE F1}, which compares two binary time-series representations, to assess whether a meaningful manually set threshold on affect data, such as the midpoint of the scale, would indicate a stronger relationship with memory labels.

\subsubsection{Euclidean Distance}
The PATE metric assesses relationships based on discrete categories rather than continuous values. This binary nature can lead to limitations when examining more nuanced or gradual changes in affect and memorability traces, as it does not account for the varying degrees of response or activation that may occur outside of strict categorical boundaries. Therefore, to quantify the relationship between emotion and memorability annotations without threshold assumptions, we use \textbf{Euclidean distance} as an additional metric. Euclidean distance offers a straightforward estimate of how similar the two time-series are to each other.

\subsubsection{Dynamic Time Warping (DTW)}
Euclidean distance does not take into account small shifts or stretches in the data, for example, if the affect signal increases consistently some time before/after the interval becomes memorable, which is plausible based on the literature \cite{Dunsmore2022SalientAttractMemory, Jeunehomme_2020}. This is why our final metric is \textbf{Dynamic Time Warping distance} (DTW) \cite{Müller2007DTW}.  DTW is a technique used to measure similarity between two temporal sequences that may vary in speed or timing. By warping the time axis, DTW aligns sequences in a flexible manner, allowing for the comparison of patterns that may be out of phase. This adaptability makes DTW particularly suitable for analysing time series data in the context of memory and affect, as it can capture meaningful correlations even when events occur at slightly different times, thus providing a more accurate representation of the underlying relationships.

Since all these metrics denote different types of relationships, we compute all four measures - PATE F1 on binary memory labels (Section \ref{memory_labels}) and binarised affect with manually set thresholds (Section \ref{affect_labels}), PATE on binary memory and continuos affect, Euclidean distance and DTW distance on continuous memory (Section \ref{memory_labels}) and continuos affect (Section \ref{affect_labels}).

Since our distance measures, such as Dynamic Time Warping (DTW) and Euclidean distance, assess differences between data points at each timestep, it is crucial to ensure that all variables are on comparable scales. The affective dimensions in our data have varying ranges (0 to 8, -8 to 8, and 0 to 11.3), while memory annotations are scaled from 0 to 1. To address this disparity, we normalised all affective dimensions to a common range of 0 to 1, ensuring consistency and meaningful comparisons across metrics.

\subsection{Statistical testing procedure\label{comparison_method}}
A crucial challenge testing the relationship between our memory and affect annotations lies in the absence of a baseline to interpret the strength of that relationship. 

Simply put, while we can compute correlation and similarity metrics (PATE F1, PATE, Euclidean distance, and DTW) to quantify the association between these variables, interpreting these values without a frame of reference leaves us uncertain about whether the observed relationships are relatively strong, weak, or even meaningful.

\begin{figure}
    \centering
    \includegraphics[width=0.8\linewidth]
    {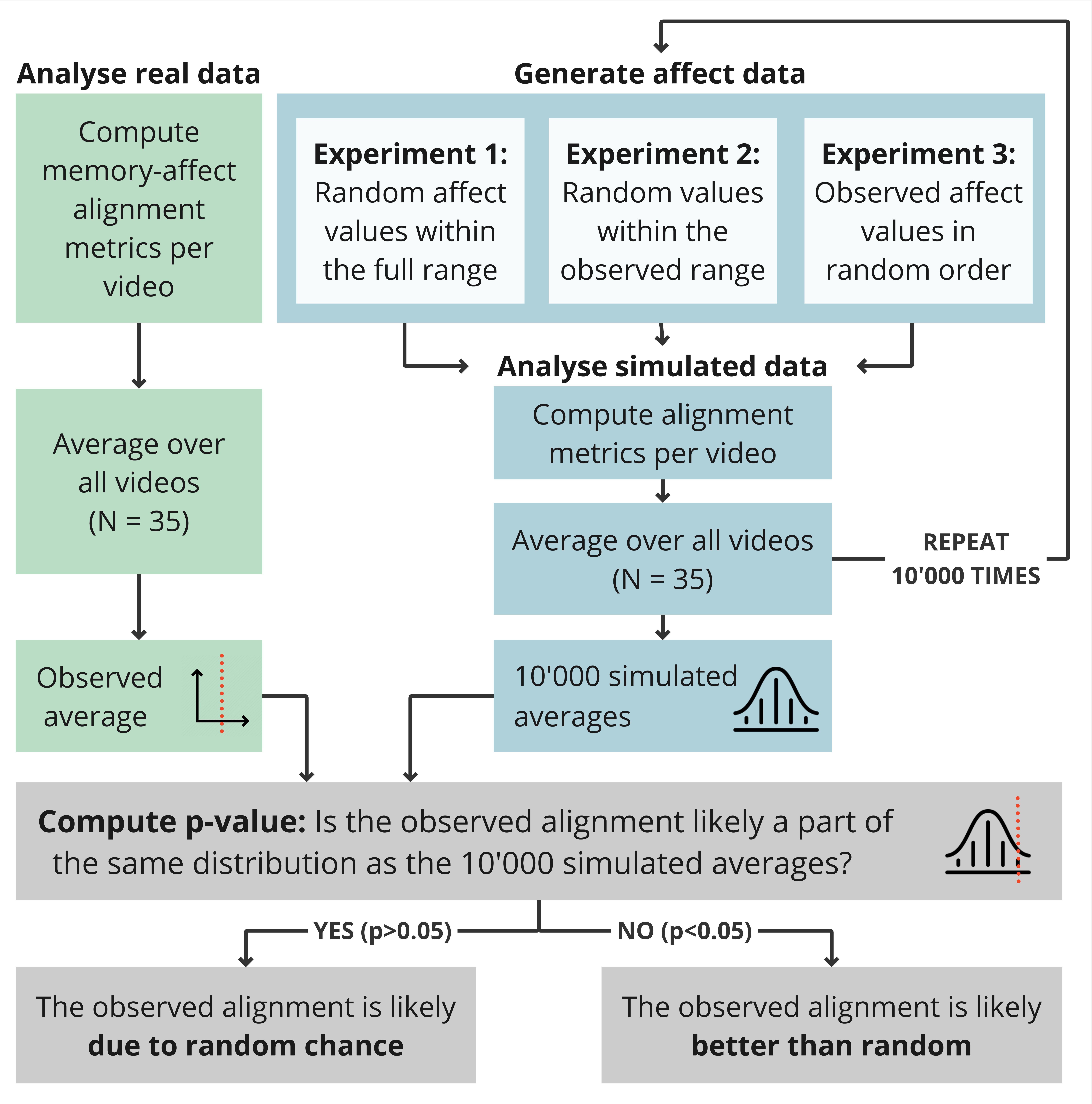}
    \caption{Illustration of the comparison procedure \label{procedure_fig}}
\end{figure}

To address this, we followed the procedure shown in Figure \ref{procedure_fig}, generating synthetic affect data with different assumptions in three experiments. These synthesised datasets act as estimates of the sampling distribution under the null hypothesis, enabling us to create a comparative framework for understanding the relationship between memory and group affect annotations. In each experiment, we simulated synthetic data with a specific hypothesis in mind: for example, assuming random affect annotations independent of memory (experiment 1), or annotations shuffled over time (experiment 2).

Using the same metrics as for the actual data, we then evaluated the relationship between each synthetic and the real memory annotations. For each iteration of an experiment, we computed the metric averages over 35 sessions, representing each session’s memory data within the dataset and comparing it to a synthesised affect annotation. To ensure the robustness of our results and account for random variations, we repeated this process for 10000 iterations, generating new synthetic data at each iteration. This produced a distribution of 10000 average metric values under the null hypothesis for each experiment.

With this distribution, we could then assess how the actual observed relationship values between real memory and affect annotations compare to the simulated point of reference. This was computed by calculating the probability that the real data metrics belong to the same distribution as the synthetic data. Specifically, by examining how likely the real data metrics are to belong within the range of values produced by the synthetic data, we could assess whether the observed relationships in the real data are likely to have occurred by random chance. If the real data metrics deviate significantly from the synthetic distribution, one would reject the null hypothesis that the association between these two variables is not different from random chance. In other words, this would suggest a meaningful difference in the alignment between group affect annotations and memory in the actual dataset compared to what we would expect under random conditions. 

The three computational experiments differed in the way the data under the null hypothesis was generated. In the 1st experiment, we generate the random affect data using the least assumptions - drawing random affect values from a random distribution of the full feature range (-8 to 8 for valence, 0 to 8 for arousal and 0 to 11.3 for intensity). This experiment is meant to answer the question of how likely it is that the relationship between memory and affect annotations is completely random. In the 2nd experiment, we generate affect annotations closer to the original data by keeping the observed range of affective labels, i.e. if in one session the arousal annotations range from 1 to 6, this is the range we use for the annotation generation for that session. In the 3rd experiment, we mimic the distribution within the observed affect when generating annotations while destroying the temporal alignment between affect and memory - we take the real affect annotations and shuffle them within the time-series. This experiment is meant to test the strength of temporal alignment between affect and memory. In all the experiments, the comparisons are performed on emotional arousal, valence and intensity annotations separately.

The decision on rejection of the null hypothesis was made using the following rule: if the p-value is significant across all 3 experiments, we could reject the overall hypothesis that the relationship between memory and observed affect annotations could belong to the same distribution as the relationship between memory and random or permuted affect annotations.

\section{Empirical Investigation}
\subsection{Experiment 1: random uniform}
\subsubsection{Simulation under the null hypothesis.} In this experiment, we compare the actual observed affect to memory alignment metrics (PATE F1, PATE, Euclidean distance, and DTW) to those on random affect data with minimal assumptions (with comparison procedure described in Figure \ref{procedure_fig} and Section \ref{comparison_method}). For this, for each of the 35 videos, we simulate affect annotations randomly drawn from a uniform distribution in the range of each affect dimension scale: 0 to 8 when simulating arousal, -8 to 8 when simulating valence and 0 to 11.3 when simulating emotional intensity. To maintain the basic structure of affect annotation, the random affect label is drawn every 15 seconds. Through this simulation, we wanted to test if the relationship that we can observe in our real data is likely random. 

\subsubsection{Results.}
\renewcommand{\arraystretch}{1.5}

\begin{table*}
\caption{Mean metric values for arousal (\textbf{A}), valence (\textbf{V}), and intensity (\textbf{I}). Green cells indicate significant p-values (p$<$0.004, Bonferroni correction), while uncolored cells are insignificant (p$\le$0.004). Arrows: $\uparrow$ (higher = more aligned), $\downarrow$ (lower = more aligned). \label{tab:results_significance}}
\centering
\setlength{\tabcolsep}{4pt}
\begin{tabular}{ccclccclccclccclccc}
\hline\hline
\multirow{2}{*}{\textbf{Aff.}} & \multirow{2}{*}{\textbf{Mem.}} & \multirow{2}{*}{\textbf{Metric}} &  & \multicolumn{3}{c}{\textbf{\begin{tabular}[c]{@{}c@{}}Mean Observed Value\end{tabular}}} &  & \multicolumn{3}{c}{\textbf{\begin{tabular}[c]{@{}c@{}}Experiment 1: H$_0$ means\end{tabular}}} &  & \multicolumn{3}{c}{\textbf{\begin{tabular}[c]{@{}c@{}}Experiment 2: H$_0$ means\end{tabular}}} &  & \multicolumn{3}{c}{\textbf{\begin{tabular}[c]{@{}c@{}}Experiment 3: H$_0$ means\end{tabular}}} \\ \cline{5-7} \cline{9-11} \cline{13-15} \cline{17-19} 
 &  &  &  & \textbf{A} & \textbf{V} & \textbf{I} &  & \textbf{A} & \textbf{V} & \textbf{I} &  & \textbf{A} & \textbf{V} & \textbf{I} &  & \textbf{A} & \textbf{V} & \textbf{I} \\ \cline{1-3} \cline{5-7} \cline{9-11} \cline{13-15} \cline{17-19} 
\textbf{Bool.} & \textbf{Bool.} & \textbf{PATE (F1) $\uparrow$} &  & 0.69 & 0.67 & 0.70 &  & \cellcolor[HTML]{9AFF99}0.63 & \cellcolor[HTML]{9AFF99}0.62 & 0.69 &  & \cellcolor[HTML]{9AFF99}0.66 & \cellcolor[HTML]{9AFF99}0.66 & 0.69 &  & 0.69 & 0.68 & 0.70 \\ \cline{1-3} \cline{5-7} \cline{9-11} \cline{13-15} \cline{17-19} 
\textbf{Bool.} & \textbf{Cont.} & \textbf{PATE $\uparrow$} &  & 0.65 & 0.66 & 0.66 &  & 0.66 & 0.65 & \cellcolor[HTML]{9AFF99}0.63 &  & 0.68 & 0.68 & 0.65 &  & 0.64 & 0.64 & 0.64 \\ \cline{1-3} \cline{5-7} \cline{9-11} \cline{13-15} \cline{17-19} 
\textbf{Cont.} & \textbf{Cont.} & \textbf{Eucl. dist. $\downarrow$} &  & 20.7 & 19.4 & 15.9 &  & \cellcolor[HTML]{9AFF99}25.1 & \cellcolor[HTML]{9AFF99}24.6 & \cellcolor[HTML]{9AFF99}24.9 &  & \cellcolor[HTML]{9AFF99}22.9 & \cellcolor[HTML]{9AFF99}22.6 & \cellcolor[HTML]{9AFF99}18.7 &  & 20.8 & 19.5 & 15.9 \\ \cline{1-3} \cline{5-7} \cline{9-11} \cline{13-15} \cline{17-19} 
\textbf{Cont.} & \textbf{Cont.} & \textbf{DTW$\downarrow$} &  & 17.7 & 16.2 & 13.1 &  & \cellcolor[HTML]{9AFF99}18.8 & \cellcolor[HTML]{9AFF99}18.3 & \cellcolor[HTML]{9AFF99}18.4 &  & 17.7 & \cellcolor[HTML]{9AFF99}17.1 & \cellcolor[HTML]{9AFF99}13.9 &  & 17.5 & 15.9 & 12.9 \\ \hline\hline
\end{tabular}
\vspace{-1em}
\end{table*}

The comparison between the observed data and the simulated random alignment metrics is shown in Table \ref{tab:results_significance}. For arousal and valence, all the metrics except for PATE showed that the observed relationship between affect and memory is unlikely due to chance (p $<$ 0.004, which is a significance level with Bonferroni correction for 12 comparisons - 4 metrics across 3 affect dimensions). For intensity, all metrics except for PATE f1 were significant.

We used PATE F1 to compare binary memory annotations with affect annotations, binarised using a meaningful threshold - the middle of Likert scales for each dimension of affect. This metric showed significant results for arousal and valence and insignificant for intensity. This means that while the relationship between observed arousal/ valence and memory is unlikely due to random chance, it is more likely due to random chance in the case of intensity. This might be due to the fact that for intensity the threshold is not as meaningful as for the measures of arousal and valence, since intensity was not measured on Likert scale and therefore the threshold is not as meaningful for that dimension. Binarising intensity might remove the important trends that contain information about a moment being encoded into group memory. This is consistent with the fact that continuous PATE on intensity showed a significant result.

In contrast, the PATE metric that operated on continuous affect, and the observed relationship showed insignificant results, appearing that the observed PATE measure is likely to belong to the same distribution as the random simulated data. This was a surprising result, which might be due to the fact that PATE which operates on continuous annotation binarises the data by generating thresholds for every observed value of the data and computes the resulting metric as the area under the curve for all those binary PATE instances. What the insignificant results might point towards is that some thresholds are more meaningful than others (this is why PATE f1 returned significant results for affect dimensions with a Likert scale). This said, PATE for intensity showed a significant result.

\begin{figure*}[ht]
    \centering
    \includegraphics[width=\textwidth]{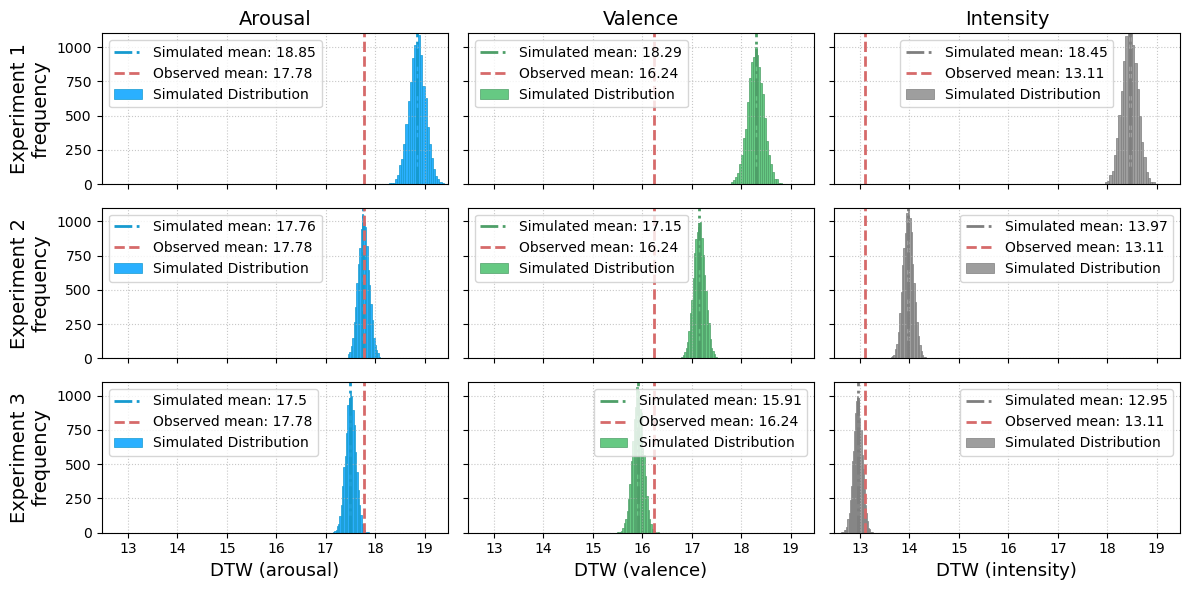}
    \caption{DTW distance results for all the experiments (each row of plots shows data for a different experiment) across the three affect dimensions - Arousal (blue), Valence (green), Intensity (gray). The Colored histogram showed the distribution of averaged DTW distance values under the null hypothesis, the red dashed line shows the averaged DTW distance value for the observed data. \textit{(Metric interpretation: The lower the DTW distance the more alignment is observed in the data.)} For other metrics, see complete figures in Appendix.}
    \label{fig:expall_DTW}
\end{figure*}

Lastly, the metrics that were computed based on continuous metrics of affect and memory, Eucledian distance and DTW distance, returned significant results across all three affect dimensions (p $\le$ 0.004). Judging by these distance metrics, whether or not the metric allows for shifts and stretches in the aligned data (in case of DTW), the observed relationship between group effect and memory is unlikely random. For illustration, Figure \ref{fig:expall_DTW} shows the difference between the observed average DTW distance and the simulated random distribution (the lower the distance the more alignment there is between the two time-series). The top row of the figure illustrates how unlikely it is that the observed relationship would be a part of a random distribution in experiment 1 (for more detail, see figures for other metrics in Appendix). 

\subsection{Experiment 2: random with observed range}
\subsubsection{Simulation under the null hypothesis.}
Experiment 2 is similar to experiment 1, but the data under the null hypothesis is simulated to be closer to the observed distribution of the affect values. This is done to ensure that the differences between the observed metrics and the random data are not due to the differences in the range of affect values represented in observed data. Specifically, experiment 1 assumes that the affect data ranges across all the possible values of a Likert scale for each affect dimension. This assumption does not hold for the real data. First, the emotion annotations across the dataset are skewed to the positive side: the arousal annotations range from 1 to 8 (with no instances of 0), the valence values range from -6 to 8 (with no instances of -7 and -8). Second, the range of captured emotions differs depending on the session, with the smallest range for arousal being 2 to 5 and the smallest range for valence being -3 to 2 for one session. Consequently, the full emotional range assumption in Experiment 1 does not hold for the data, which might exaggerate the differences between the observed and simulated data. Instead, in experiment 2 we simulate the data based on the affect values represented in each video. Similarly to experiment 1, random values are drawn for each 15 seconds of each video, but they are drawn from the set of values present in the affect data for that specific video. For example, if an observed annotation has arousal values [1, 3, 5, 6, 8], the simulated arousal data would draw random values from that set for every 15 seconds of the video. Except for simulation assumptions, the analysis is done according to the same procedure as experiment 1 (see procedure details in Figure \ref{procedure_fig} and Section \ref{comparison_method}).

\subsubsection{Results.}
As shown in Table \ref{tab:results_significance}, the results for Experiment 2 are similar to Experiment 1, indicating that the observed relationship between group affect and memory annotations is unlikely to belong to the same distribution as the relationships between simulated affect and memory. This is true for all the values except for PATE F1 for intensity, PATE for all the dimensions, and DTW for arousal. In comparison to experiment 1, two more values became insignificant (DTW distance for arousal and PATE F1 for valence). 

\subsection{Experiment 3: temporal SHUFFLE}
\subsubsection{Simulation under the null hypothesis.}
While experiment 1 and 2 show a comparison of observed affect to memory alignment metrics to completely random data, experiment 3 investigates the importance of temporal alignment between the affect and memory annotations. This means that in comparison to experiments 1 and 2, the simulated data in Experiment 3 represents the exact distribution of affect values in the observed data, but in randomised order within each video. In other words, we simulate data by shuffling the actual affect annotations for each video, destroying the temporal alignment between memory and affect labels.  Similar to experiments 1 and 2, we are keeping the integrity of the basic data structure by shuffling windows of 15 seconds within each video. Apart from this, the analysis procedure is the same as for experiments 1 and 2 (see procedure details in Figure \ref{procedure_fig} and Section \ref{comparison_method}).
\subsubsection{Results.}
Surprisingly, all the metrics across all the affect dimensions showed an insignificant difference from the simulated distribution in Experiment 3 (see Table \ref{tab:results_significance}). This means that under the assumptions of this experiment, we cannot reject the null hypothesis, meaning that the observed relationship between group affect and memory annotation is likely to belong to the same distribution as the memory compared with temporally shuffled affect data.

\section{Discussion}
We conducted three computational experiments to evaluate the relationship between memory and group affect annotations.
Despite significant values in the first two experiments, we can conclude that the null hypothesis cannot be rejected, since there are no metrics or affect dimensions for which all 3 experiments produced a significant p-value (see our null hypothesis rejection rule in Section \ref{comparison_method}). 

\textbf{Annotation Perspectives.}
One of the core assumptions motivating this study was that emotional experiences influence memory encoding, a well-established link in cognitive science \cite{Dolcos_2004_EmotionAmygdaleMemory, Shohamy_2010_MotRelevenceInfMemory}. However, in affective computing, emotional states are often inferred from third-party annotations of observed behaviour rather than first-party reports of experienced emotions \cite{AffectDetection2015}. Our findings indicate that this distinction is crucial: while experienced emotions are directly tied to personal relevance and cognitive appraisal \cite{Lewis2005Appraisal, Scherer_1984}, third-party affect annotations reflect an external interpretation of group behaviour that may not reliably map onto internal memory processes. Experiment 3 has demonstrated that the significant effects seen in Experiments 1 and 2 were likely due to the differences in distributions of affect labels in the observed data, rather than that affect annotations align with memory annotations better than chance. This discrepancy aligns with potential concerns regarding third-party annotations, which are known to be influenced by external factors such as social norms and individual expressivity \cite{Cameron2018Suppression}. For instance, emotional behaviours may not always correspond to experienced emotions due to social masking, such as hiding frustration with a polite smile. 

\textbf{Continuous Conceptualisation.}
A second key issue is how affect and memory are conceptualised over time. To the best of our knowledge, traditional memory studies assess emotional experience and memorability as static states, typically using retrospective self-reports \cite{Talarico2004}. In contrast, MER applications require continuous, time-aligned affective labels to enable real-time system responses \cite{mariooryad2015a}. By testing the relationship between time-continuous affect annotations and memory, we examined whether existing findings on affect and memory translate to a dynamic, multimodal annotation setting. While time-continuous annotations are widely used in affective computing to capture fine-grained emotional fluctuations \cite{zhang2020a}, previous research linking affect and memory has not employed such temporally granular methods. Our results suggest that while emotional intensity, valence, and arousal may contribute to memory encoding, their influence is not reliably captured through continuous third-party annotations. This aligns with previous findings that retrospective memory reports are influenced by post-event reconstruction biases, which are not accounted for in continuous affect or memory annotation frameworks \cite{mcgaugh2013, dudzik2023}. The lack of a robust relationship between continuously observed affect and memory shown by our results suggests that real-time affective annotations alone may not be sufficient for predicting memory in conversational contexts, necessitating alternative methodologies that consider retrospective appraisal effects and the temporal structure of memory retrieval.

\textbf{Group-Level Analysis.}
Finally, this study contributes to the growing need for group-based emotions and memory research. While prior work has examined emotion’s role in individual memory encoding \cite{Sharot2004, Cahill_1996_MemoryModulation, Maljkovic_Martini_2005ValenceAffectsMemory}, our study explicitly considers group dynamics, a key factor in real-world settings like meetings and collaborative tasks \cite{Samiha2021MeetingCoach, Trian2019GroupAffectForLearning, Järvenoja2017GroupEmotionsForCollaboration}. Our results suggest that group emotion annotations, which capture collective emotional states rather than individual experiences, may fail to account for memorability. This could be due to the fact that group memorability annotations might not capture an emergent group-level processes in a way that group affect does \cite{Smith2007TrulyGroupAffect}, since they are aggregated from individual memory reports rather than inferred from group states to begin with. Another possible reason is that people express emotions differently in group settings compared to one-on-one conversations or non-social situations. For example, research suggests that emotions tend to be expressed more strongly in dyads than in larger groups \cite{groupsVSdyads}. Additionally, group members often adjust their emotional expressions to match each other, a phenomenon known as emotional convergence \cite{BarsadeGibson2012GroupAffect, Yoon2013CohesionHigherInGroups}
This convergence may dilute individual emotional expressions, driving it further from the individual experienced emotion that would have been connected to memorability.

\section{Conclusions}
This study investigated the potential of using group emotion annotations as proxies for conversational memorability in multi-party settings. By analysing the relationship between memorability and affective dimensions using data from the MeMo corpus \cite{Tsfasman2024}, we conducted a series of computational experiments comparing affect annotations, particularly arousal, valence, and intensity and memorability. 

While the relationship between affect and memory showed to be significantly different than random (see experiment 1), when correcting for distribution biases of real affect annotations, experiments showed that such a relationship is insignificantly different than random (experiment 2) or temporally shuffled data (experiment 3). Overall, our analyses revealed that the observed metrics for real data did not deviate meaningfully from the distributions derived from synthetic data generated under null hypotheses. This finding suggests that, within the scope of this dataset and methodology, affect annotations (in terms of arousal, valence, or intensity) do not serve as reliable proxies for conversational memorability. 
Therefore, despite a common belief that affective states capture inter-personal relevance in alignment to memory, our findings highlight the need for dedicated research on modelling memorability - a distinct indicator of long-term event relevance.

While emotions and memory have been conceptually linked in cognitive science, our findings suggest that this relationship may not translate to the settings typical to Affective Computing applications. These applications traditionally operate with continuos affect annotations, collected from third party observers, that rely on participants' behaviour to infer their affective states (illustrated in Figure \ref{fig:study_object}). In contrast, prior research on the emotion-memory link has used static self-reports or physiological measures of emotional experience. Our study shows that observed affect annotations, particularly at the group level, do not meaningfully align with memorability, highlighting the importance of these conceptual differences between the operationalised constructs. 
Although the memory-emotion link has been treated as a given in some Intelligent Systems applications \cite{kasap_interacting_2010, Ahmad2021EmotionMemoryModel, Moerland2017EmotionForRL}, we urge future research to account for differences in how these constructs are defined and measured (e.g., third-party vs. first-party, group vs. individual level, continuous vs. static). Failing to consider these discrepancies may lead to inaccurate transfers of empirical findings into computational applications. 

To better understand the emotion-memorability link, we recommend further research into whether individual-level perceived affect annotations exhibit the same lack of relationship with individual memorability as observed at the group level. We also suggest examining these phenomena in face-to-face interactions rather than online video conferencing to determine whether the setting influences the relationship between memory and affect annotations. Lastly, there may be differences in the types of memorable moments that are linked to affect and those that are not, warranting further investigation into the contextual factors shaping memory and affect.

\vspace{-1.5em}
\small
\bibliographystyle{IEEEtran}  
\bibliography{main.bib}  


\onecolumn
\setcounter{section}{0}
\setcounter{table}{0}
\newpage
\section*{Appendix 1: Simulation procedure}
\subsection*{Experiment 1} In this experiment, we compare the actual observed affect to memory alignment metrics (PATE F1, PATE, Euclidean distance, and DTW) to those on random affect data with minimal assumptions. For this, for each of the 35 videos, we simulate affect annotations randomly drawn from a uniform distribution in the range of each affect dimension scale: 0 to 8 when simulating arousal, -8 to 8 when simulating valence and 0 to 11.3 when simulating emotional intensity. To maintain the basic structure of affect annotation, the random affect label is drawn every 15 seconds. Through this simulation, we wanted to test if the relationship that we can observe in our real data is likely due to absolute chance. 

\subsection*{Experiment 2} Experiment 2 is similar to Experiment 1, but with null hypothesis data simulated to match the observed distribution of affect values. This ensures that differences between the observed and random data are not due to discrepancies in the range of affect values. In Experiment 1, the affect data was assumed to span the entire Likert scale, but this does not hold for the real data, with emotion annotations positively skewed: arousal ranges from 1 to 8 (with no 0s), and valence ranges from -6 to 8 (with no -7 or -8). Additionally, the range of emotions varies by session, with some sessions showing smaller ranges for arousal (2 to 5) and valence (-3 to 2). To address this, Experiment 2 simulates data based on the affect values present in each video. Random values are drawn for each 15-second interval, but only from the set of values observed in that specific video. For example, if the arousal annotations include [1, 3, 5, 6, 8], simulated data for arousal would randomly select values from this set for every 15 seconds. The analysis follows the same procedure as Experiment 1.

\subsection*{Experiment 3}
While experiment 1 and 2 show a comparison of observed affect to memory alignment metrics to completely random data, experiment 3 investigates the importance of temporal alignment between the affect and memory annotations. This means that in comparison to experiments 1 and 2, the simulated data in Experiment 3 represents the exact distribution of affect values in the observed data, but in randomised order within each video. In other words, we simulate data by shuffling the actual affect annotations for each video, destroying the temporal alignment between memory and affect labels.  Similar to experiments 1 and 2, we are keeping the integrity of the basic data structure by shuffling windows of 15 seconds within each video. Apart from this, the analysis procedure is the same as for experiments 1 and 2.

\clearpage
\section*{Appendix 2: Visualisation of metric comparison for all experiments}

\begin{figure*}[ht]
    \centering
    \includegraphics[width=\textwidth]{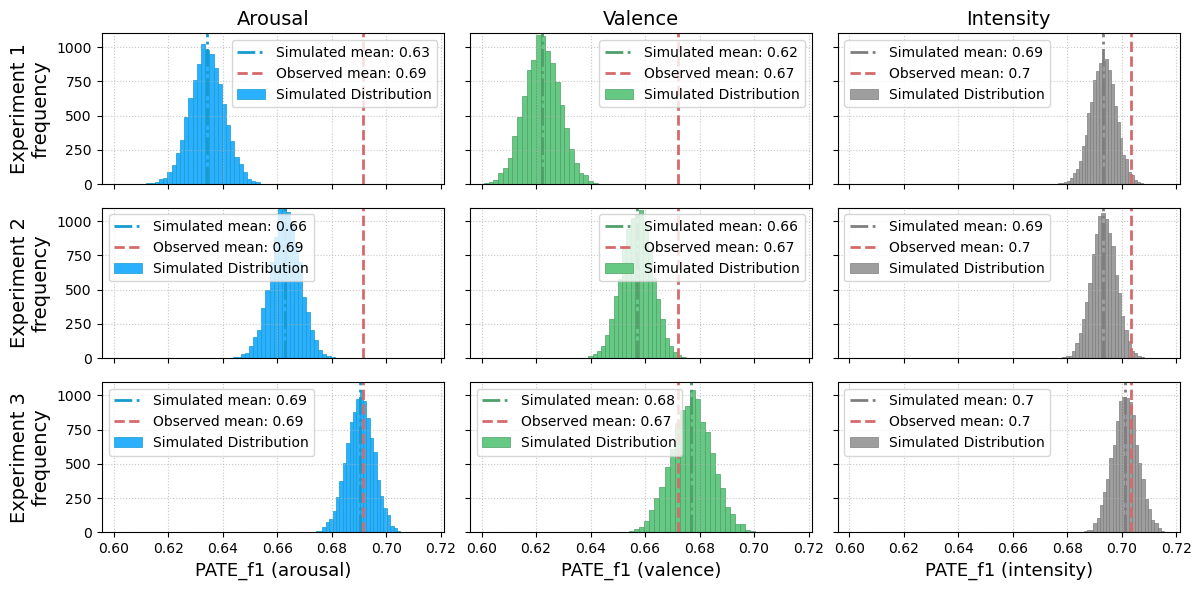}
    \caption{\textbf{PATE F1} results for all the experiments (each row of plots shows data for a different experiment) across the three affect dimensions - Arousal (blue), Valence (green), and Intensity (grey). The Colored histogram showed the distribution of averaged PATE F1 values under the null hypothesis, the red dashed line shows the averaged PATE value for the observed data. \textit{(Metric interpretation: The higher PATE F1 the more alignment is observed in the data.)}}
    \label{fig:expall_PATEF1}
\end{figure*}
\clearpage
\subsection*{PATE}
\begin{figure*}[ht]
    \centering
    \includegraphics[width=\textwidth]{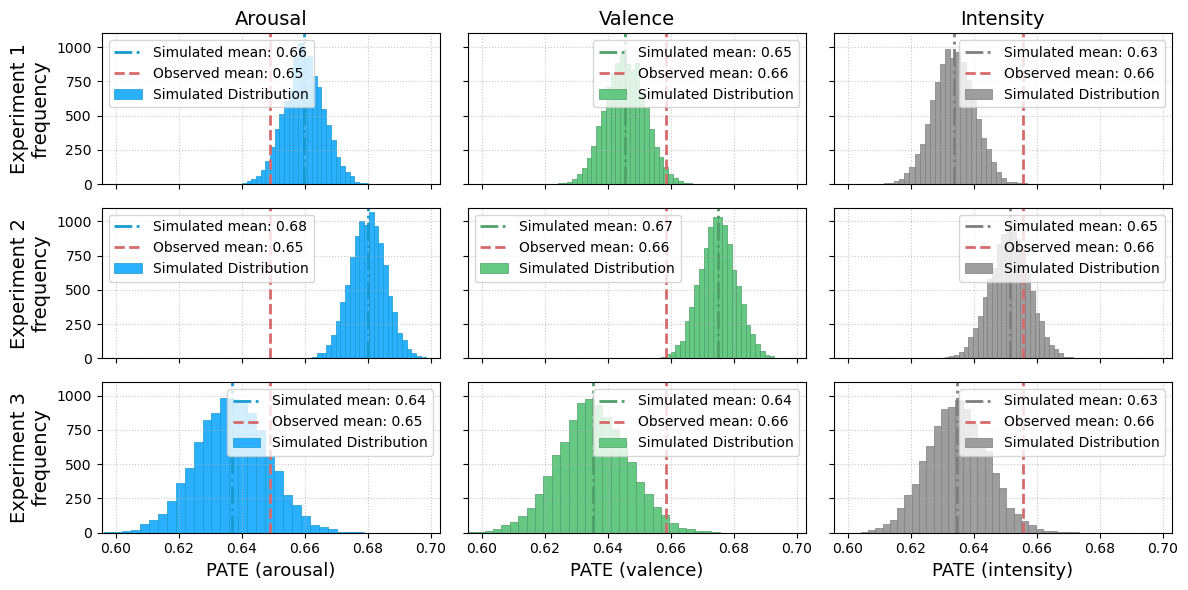}
    \caption{\textbf{PATE} results for all the experiments (each row of plots shows data for a different experiment) across the three affect dimensions - Arousal (blue), Valence (green), Intensity (grey). The Colored histogram showed the distribution of averaged PATE values under the null hypothesis, and the red dashed line shows the averaged PATE value for the observed data. \textit{(Metric interpretation: The higher PATE the more alignment is observed in the data.)}}
    \label{fig:expall_PATE}
\end{figure*}
\clearpage
\subsection*{Eucledian Distance}
\begin{figure*}[ht]
    \centering
    \includegraphics[width=\textwidth]{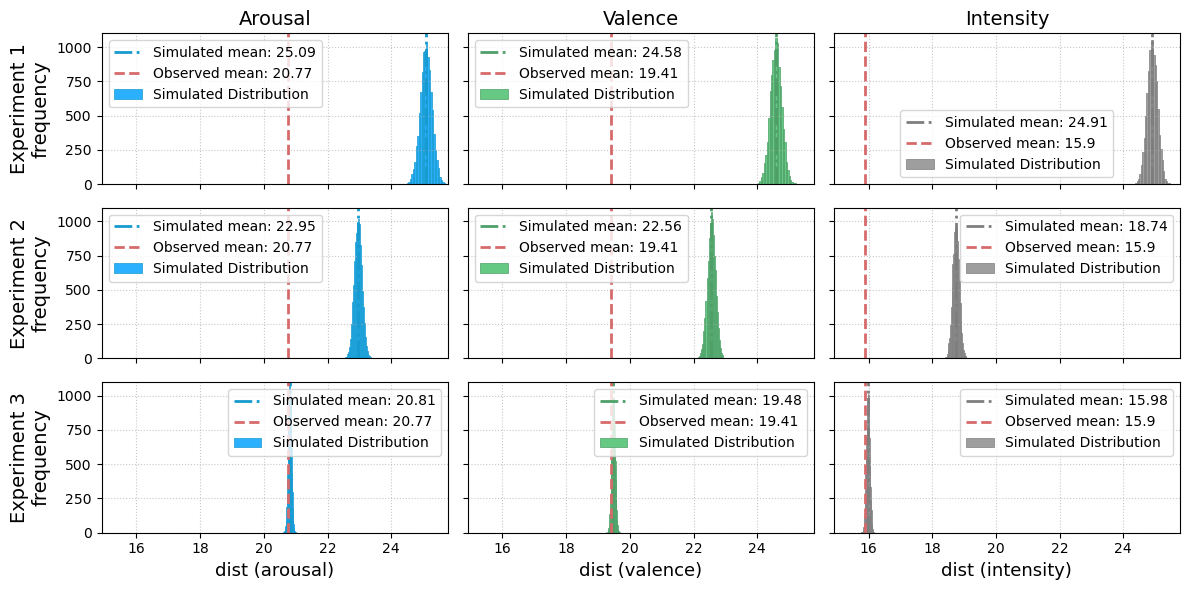}
    \caption{\textbf{Eucledian distance} results for all the experiments (each row of plots shows data for a different experiment) across the three affect dimensions - Arousal (blue), Valence (green), Intensity (grey). The Colored histogram showed the distribution of averaged Euclidean distance values under the null hypothesis, the red dashed line shows the averaged Euclidean distance value for the observed data. \textit{(Metric interpretation: The lower the Euclidean distance the more alignment is observed in the data.)}}
    \label{fig:expall_dist}
\end{figure*}

\vfill




\vspace{-1.5em}
\section*{Author Biographies}
\vspace{-4em}
\begin{IEEEbiography}
[{\includegraphics[width=1in,height=1.75in,clip,keepaspectratio]{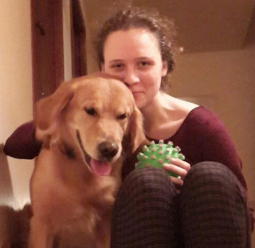}}]{Maria Tsfasman} (a.k.a Masha) Her interdisciplinary memory research aims to enhance machine understanding of humans and gain insights into human cognition. She holds a BSc in Linguistics from HSE University, Moscow, and a MSc in AI with distinction from Radboud University, Nijmegen. Before her PhD, she is proud to have been a research assistant at ISIR, Sorbonne University, and IRCN, University of Tokyo. 
\end{IEEEbiography}
\vspace{-4em}

\begin{IEEEbiography}
[{\includegraphics[width=1in,height=1.75in,clip,keepaspectratio]{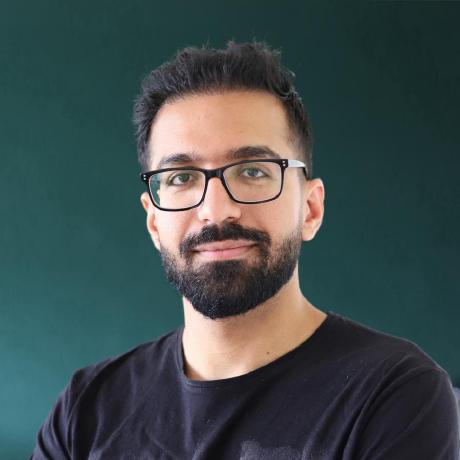}}]{Ramin Ghorbani} received the B.S. degree in industrial engineering from Yazd University, Iran, in 2017, and the M.S. degree in system optimisation (data mining in healthcare) from the Iran University of Science and Technology, Iran, in 2019. Since 2018, he has been a Research Assistant. His research interests include artificial intelligence, machine learning, and data analysis, with a passion for health informatics and bioinformatics.
\end{IEEEbiography}
\vspace{-4em}

\begin{IEEEbiography}
[{\includegraphics[width=1in,height=1.75in,clip,keepaspectratio]{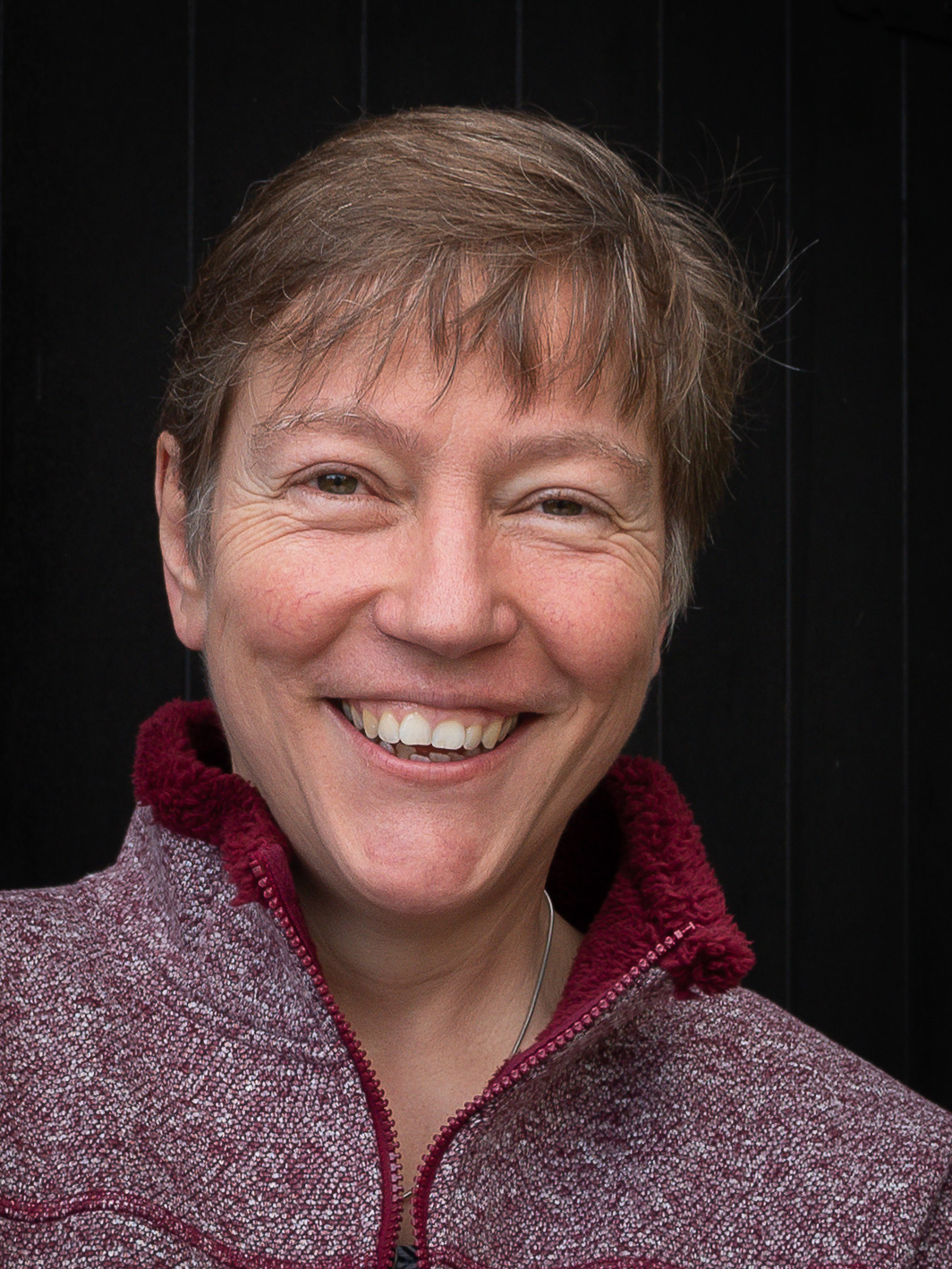}}]{Catholijn M. Jonker} researches hybrid intelligence and interactive intelligent processes such as strategic decision making, negotiation, teamwork and the design of agent technology. She is board member of IFAAMAS, and vice-coordinator of the Hybrid Intelligence Centre. She is a member of the Royal Holland Society of Sciences and Humanities, a Fellow of EurAI, and member of the Academia Europaea. She is a co-founding member of the Netherlands Academy of Engineers. 
\end{IEEEbiography}
\vspace{-3em}

\begin{IEEEbiography}
[{\includegraphics[width=1in,height=1.75in,clip,keepaspectratio]{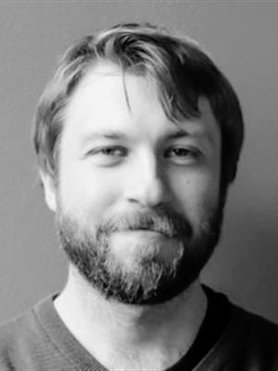}}]{Bernd Dudzik} is an Assistant Professor in the Department of Intelligent Systems at \textit{TU Delft}. His research interests focus on Affective Computing and AI-Human collaboration. In particular, his work explores context-sensitive approaches for multimodal modelling of human cognitive-affective processes (e.g., memory recollection or cognitive appraisals). Bernd is an active member of the AAAC and Associate Editor for the IEEE Transactions on Affective Computing. \end{IEEEbiography}
\vspace{-4em}
\vfill
\end{document}